\documentclass[%
 aip,
 amsmath,amssymb,
 reprint,%
]{revtex4-1}

\usepackage{graphicx}
\usepackage{dcolumn}
\usepackage{bm}

\usepackage[utf8]{inputenc}
\usepackage[T1]{fontenc}
\usepackage{mathptmx}
\usepackage{etoolbox}

\makeatletter
\def\@email#1#2{%
 \endgroup
 \patchcmd{\titleblock@produce}
  {\frontmatter@RRAPformat}
  {\frontmatter@RRAPformat{\produce@RRAP{*#1\href{mailto:#2}{#2}}}\frontmatter@RRAPformat}
  {}{}
}%
\makeatother
\begin{document}


\title[Surrogate models to optimize PEALD in high aspect ratio features]{
Surrogate models to optimize plasma assisted atomic layer deposition
in high aspect ratio features}
\author{Angel Yanguas-Gil}
\email{ayg@anl.gov}
\author{Jeffrey W. Elam}%
\affiliation{ 
Applied Materials Division, Argonne National Laboratory,
Lemont, IL 60439 (USA)
}

\date{\today}

\begin{abstract}
In this work we explore surrogate models to optimize plasma enhanced atomic layer deposition (PEALD) in high aspect ratio features. In plasma-based processes such as PEALD and atomic layer etching, surface recombination can dominate the reactivity of plasma species with the surface, which can lead to unfeasibly long exposure times to achieve full conformality inside nanostructures like high aspect ratio vias. Using a synthetic dataset based on simulations of PEALD, we train artificial neural networks to predict saturation times based on cross section thickness data obtained for partially coated conditions. The results obtained show that just two experiments in undersaturated conditions contain enough information to predict saturation times within 10\% of the ground truth. A surrogate model trained to determine whether surface recombination dominates the plasma-surface interactions in a PEALD process achieves 99\% accuracy. This demonstrates that machine learning can provide a new pathway to accelerate the optimization of PEALD processes in areas such as microelectronics. Our approach can be easily extended to atomic layer etching and more complex structures. 
\end{abstract}

\maketitle

\section{Introduction}

Plasma enhanced self-limited techniques such as atomic layer deposition (ALD) and plasma-assisted atomic layer etching (ALE) are key tools in semiconductor processing\cite{Profijt_PEALDreview_2011,Knoops_PEALDreview_2019,Kanarik_ALE_review_2015, Lill_ALE_2022}. They both rely on the use of plasmas as coreactants to a metalorganic or halide precursor, leveraging highly reactive plasma species to complete an ALD or ALE cycle and eliminate any remaining surface ligands from the surface. To accomplish this, ALD and ALE require separate exposures to the precursor and the plasma to ensure that each reaction is taken to full completion. Determining the magnitude of these exposure times, typically referred to as dose times, is a crucial part of process optimization in ALD and ALE.

One of the challenges of plasma based processes is the presence of multiple surface reaction pathways: in addition to the precursor-surface interactions conducive to growth or etching, surface recombination can play a key role in the ALD or ALE of high surface area substrates, such as deep trenches often found in microelectronics (Figure 1). When surface recombination is dominant, it can lead to exponentially long increases in the required dose times with increasing aspect ratios\cite{Knoops_conformality_2010,YanguasGil_conformalALD_2012, Arts_recombination_2021}. One of the challenges of exploring new processes is that the surface kinetics is usually poorly understood.

Quickly identifying the optimal exposures required to homogeneously coat (or etch) a given substrate is paramount to determine the practical feasibility of a process. This task is particularly challenging if the substrate contains high aspect ratio features. A common way of verifying the conformality of a process is through the inspection of cross section images of the high aspect ratio substrates using either scanning or transmission electron microscopy. There is an opportunity to leverage the information contained in cross section profiles for undersaturated conditions (i.e. Figure 1) to help drive the optimization process. We can envision two different scenarios of practical relevance: in a first case, we wish to identify optimal conditions for the substrate under test. In a second scenario, we would like to identify the optimal conditions to a different substrate. This can be helpful, for instance, when transferring a process from R\&D to manufacturing. In both cases, we would like to do so in as few experiments as possible.

\begin{figure}
\includegraphics[width=7.5cm]{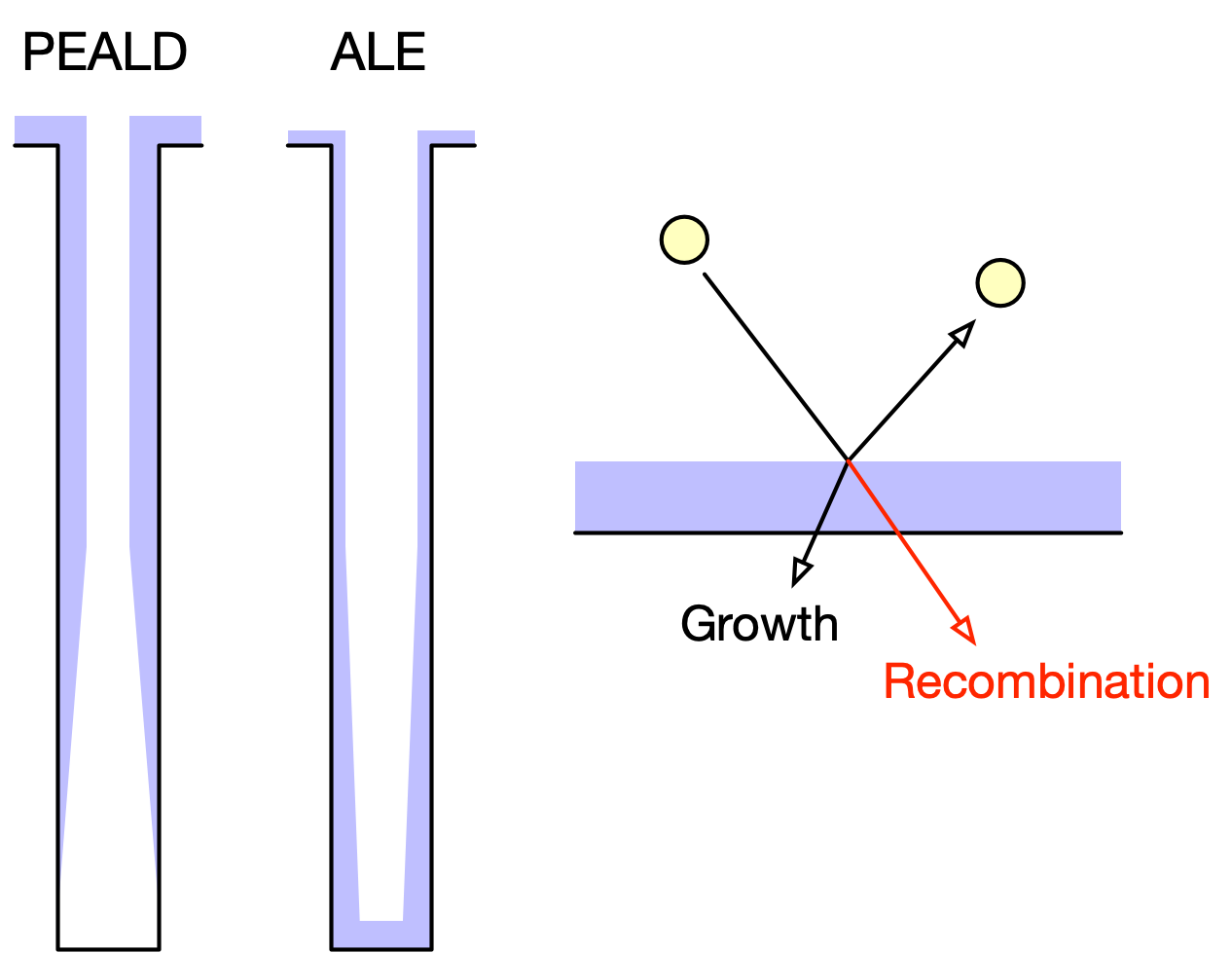}
\caption{\label{fig1} Cross sectional images of high aspect ratio features are used to validate the conformality during growth or etching. In the case of plasma-assisted processes, surface recombination of plasma species can pose a challenge to achieve conformal growth or etching within high aspect ratio and high surface area materials.}
\end{figure}

Traditional approaches to tackle this problem have relied on simulations to extract the kinetic parameters by fitting the models to the experimental data\cite{Arts_recombination_2019, Arts_TMAsticking_2019}. An alternative approach is the development of surrogate models that can directly predict the optimal dose time directly from experimental growth profiles in undersaturated conditions. An important advantage of this approach is that, if these surrogate models are trained on a broad enough set of conditions, they can predict optimal conditions for any process without the need of knowing any additional information about the underlying mechanisms or running additional simulations. This is something that we have observed in thermal ALD processes, where a single snapshot of the thickness inhomogeneity in an underdosed process can be used to accurately predict the minimum exposure required to achieve homogeneous growth across the whole substrate\cite{YanguasGil_ALDML_2022}. 

This work focuses on exploring the potential of surrogate models for optimization of plasma-assisted ALD processes based on cross section data. Questions that will be addressed include: 1) is it possible? 2) how many independent experiments are needed? 3) how many measurements would we need? To answer these questions, we have generated a synthetic dataset focused on circular vias, but the same approach could be applicable to other types of structures common in microelectronics.

\section{Methodology}

\subsection{Dataset}

To explore the optimization of exposure times in plasma-assisted ALD, we have created a dataset comprising thickness values determined at different depths within high aspect ratio features. In a prior work we demonstrated that simulations can play a useful role in generating the data required to train surrogate models\cite{YanguasGil_ALDML_2022}. Therefore, we have used a model for reactive transport inside high aspect ratio features that has been previously shown to match well with experimental data\cite{Arts_TMAsticking_2019}. Details about the model can be found elsewhere\cite{YanguasGil_conformalALD_2012}.

Each entry in the dataset corresponds to a different self-saturated process, where the underlying surface kinetics and is randomly chosen to provide a wide range of possible plasma-assisted ALD processes. The surface kinetics is characterized by a sticking probability for the self-limited process, $\beta_s$, and a recombination probability for the plasma species with the surface, $\beta_r$. These parameters were varied between 10$^{-1.5}$ to 10$^{-4.5}$ and 10$^{-2}$ and 10$^{-6}$, respectively. Our dataset contains thickness profiles for two different dose times. The first dose time $t_1$ was selected so that it would lead to an average thin film thickness inside the high aspect ratio feature between 20\% to 30\% of a fully saturated, conformal process. The specific value was sampled from a uniform distribution between these two values.  The second exposure time was twice as long: $t_2=2\,t_1$. Finally, for each entry we computed the total saturation time $t_\mathrm{sat}$ required to achieve 99\% step coverage.

Predicting actual times requires considering additional process parameters, such as the precursor partial pressure. If instead we focus on predicting the multiplying factor that we need to apply to achieve full saturation, so that 
$t_\mathrm{sat} = \kappa\, t_1$, we do not need to consider any additional parameters. The problem therefore reduces to predicting the multiplying factor based on the thickness measurements along the high aspect ratio feature:
\begin{equation}
    (\bf{x}_1, \bf{x}_2) \rightarrow \kappa
\end{equation}
Here $\bf{x}_1$, $\bf{x}_2$ are the thickness values obtained for the two dose times $t_1$ and $t_2$. In addition to the saturation dose time, we have also stored the ALD sticking probability, $\beta_s$, and the recombination probability, $\beta_r$, for each of the processes. This will be used to evaluate the surrogate model's ability to extract kinetic data from thickness profiles.
The complete dataset comprises 20 thickness measurements at equally spaced positions inside the feature. Other datasets for 10, 8, 5, and 4 points were obtained by sampling points from the denser dataset. In this work we considered a circular via with an aspect ratio of 50. The simulation can be easily generalized to consider other aspect ratio or other types of structures such as trenches or tapered vias. We considered a total of 25,000 independent processes: 20,000 of those were used for training, and the remaining 5,000 were used for testing. The predictive capabilities of the surrogate model are therefore evaluated for profiles the model has not seen during training.

\begin{figure}
\includegraphics[width=7.5cm]{ 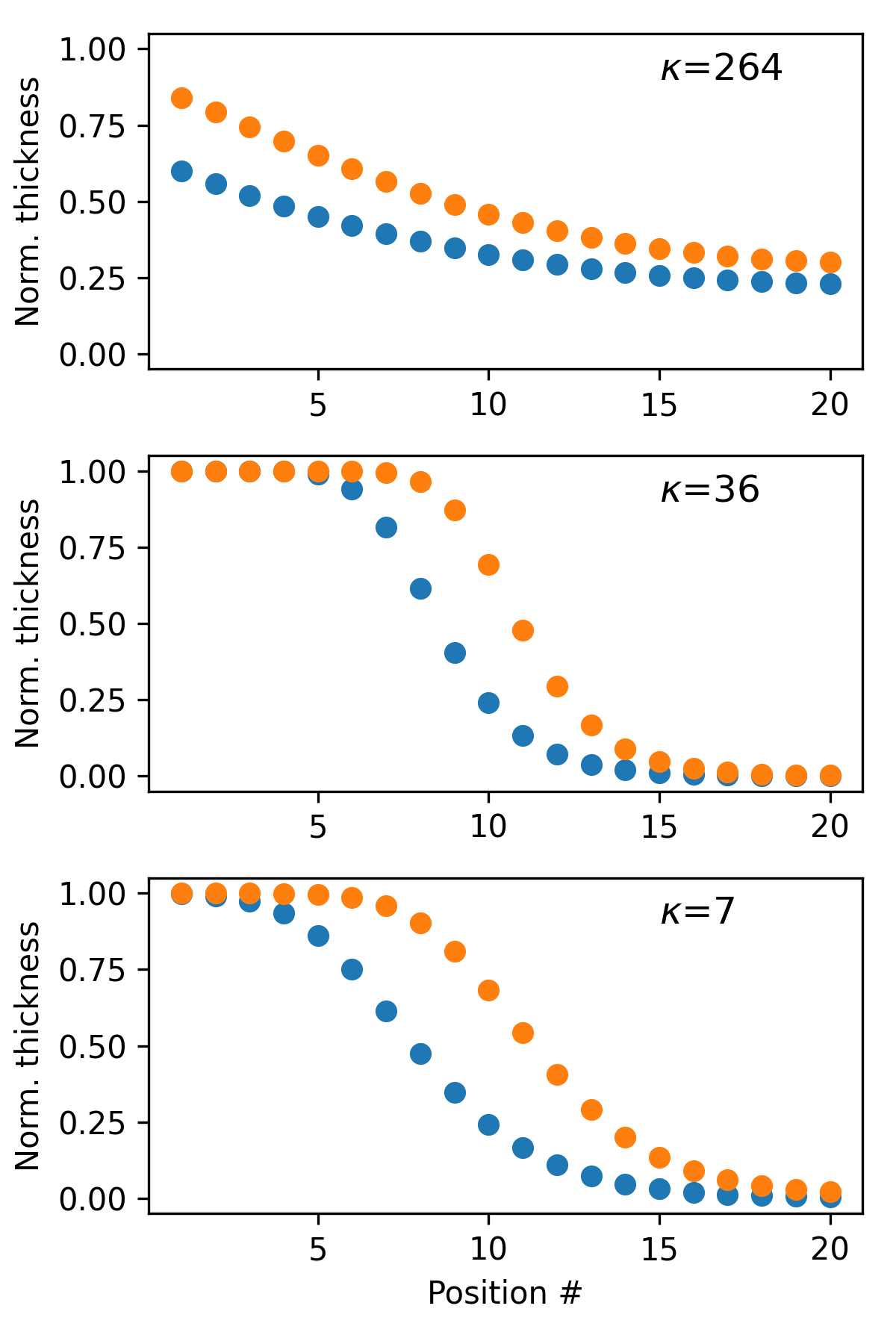}
\caption{\label{fig2} Three snapshots of normalized thickness as a function of depth for PEALD processes. These are obtained for two experiments with increasing exposure times in a circular via of aspect ratio 50. For each plot we also show the normalized saturation time $\kappa$, defined as the multiplier that we need to apply to the shorter dose time in the experiment to achieve full saturation within the high aspect ratio feature. The sticking and recombination probabilities are:
(A)  $\beta_s=3.9\times 10^{-5}$, $\beta_r=3.2\times 10^{-3}$; 
(B) $\beta_s=2.4\times 10^{-2}$, $\beta_r=3.1\times 10^{-3}$; 
(C) $\beta_s=1.1\times 10^{-2}$, $\beta_r=6.4\times 10^{-6}$}
\end{figure}

In Figure \ref{fig2} we show the resulting cross sectional profiles for a few samples in the dataset. As expected, the thickness dependence with depth and the increase in thickness when the dose times are doubled varies from process to process. Each plot shows the value of the normalized saturation time $\kappa$ required to achieve full saturation in the via, which spans two orders of magnitude. These range from a factor or 3-4 for processes where surface recombination is negligible and have low sticking probabilities, to more than 1000 for processes that are recombination-limited.

\subsection{Surrogate model}

In this work we focus on surrogate models based on artificial neural networks. In particular, we seek to explore the surrogate model’s ability to learn functional relationship between the thickness profiles and the predicted saturation time required to homogeneously coat a given high aspect ratio feature.

We consider feedforward networks with two hidden layers. Input, hidden, and output layers are all linked through all-to-all connections. The inputs to each hidden layers are passed through a rectified linear unit, so that:
\begin{equation}
\bf{a}_{i+1} = \mathrm{ReLU}\left(\mathrm{W}\bf{a}_i+\bf{b}
\right)
\end{equation}

Here $\bf{a}_i$ is the input to layer $i+1$, W is the weight matrix, $\bf{b}$ is the bias vector and $\bf{a}_{i+1}$ is the output of layer $i+1$.

The surrogate models take either $N$ or $2N$ inputs depending on whether we consider a single profile or two different profiles coming from two exposure times as inputs. The models return a single value, in this case the normalized saturation time, $\kappa$, described above. The size of the two hidden layers are $M_1$ and $M_2$, respectively. In this work we explored layer sizes ranging from 10 to 50 neurons.
In addition to predicting the saturation times, we have also considered surrogate models that can extract the sticking and recombination probabilities from the thickness data using the same network architecture.

\subsection{Training and testing}

The training and testing were carried out using Pytorch, a free open source framework for deep learning\cite{pytorch}. We trained the networks against the training sets using stochastic gradient descent methods using the same procedure described in a prior work\cite{YanguasGil_ALDML_2022}. One of the challenges of the current dataset is that the multiplicative factor $\kappa$ and the sticking and recombination probabilities span three orders of magnitude. To improve its accuracy, the surrogate model is trained to predict the logarithm of each of the magnitudes. The implementation of this network and the training scripts can be found online at https://github.com/aldsim/plasmaml. 

For analysis and visualization of the surrogate model outputs we have used the relative error between the predicted normalized time and the ground truth, defined as:
\begin{equation}
\varepsilon=\frac{t_\mathrm{pred}-t_\mathrm{sat}}{t_\mathrm{sat}}
\end{equation}

When we compute this magnitude for the testing dataset, we obtain a distribution of errors. For a high performing, unbiased surrogate model, we expect the mean prediction error,
$\mu_\varepsilon$, to be close to zero, and the standard deviation, $\sigma_\varepsilon$, to be as low as possible.

\section{Results\label{sec:results}}

\subsection{Predicting normalized saturation times}

\begin{figure}
\includegraphics[width=7.5cm]{ 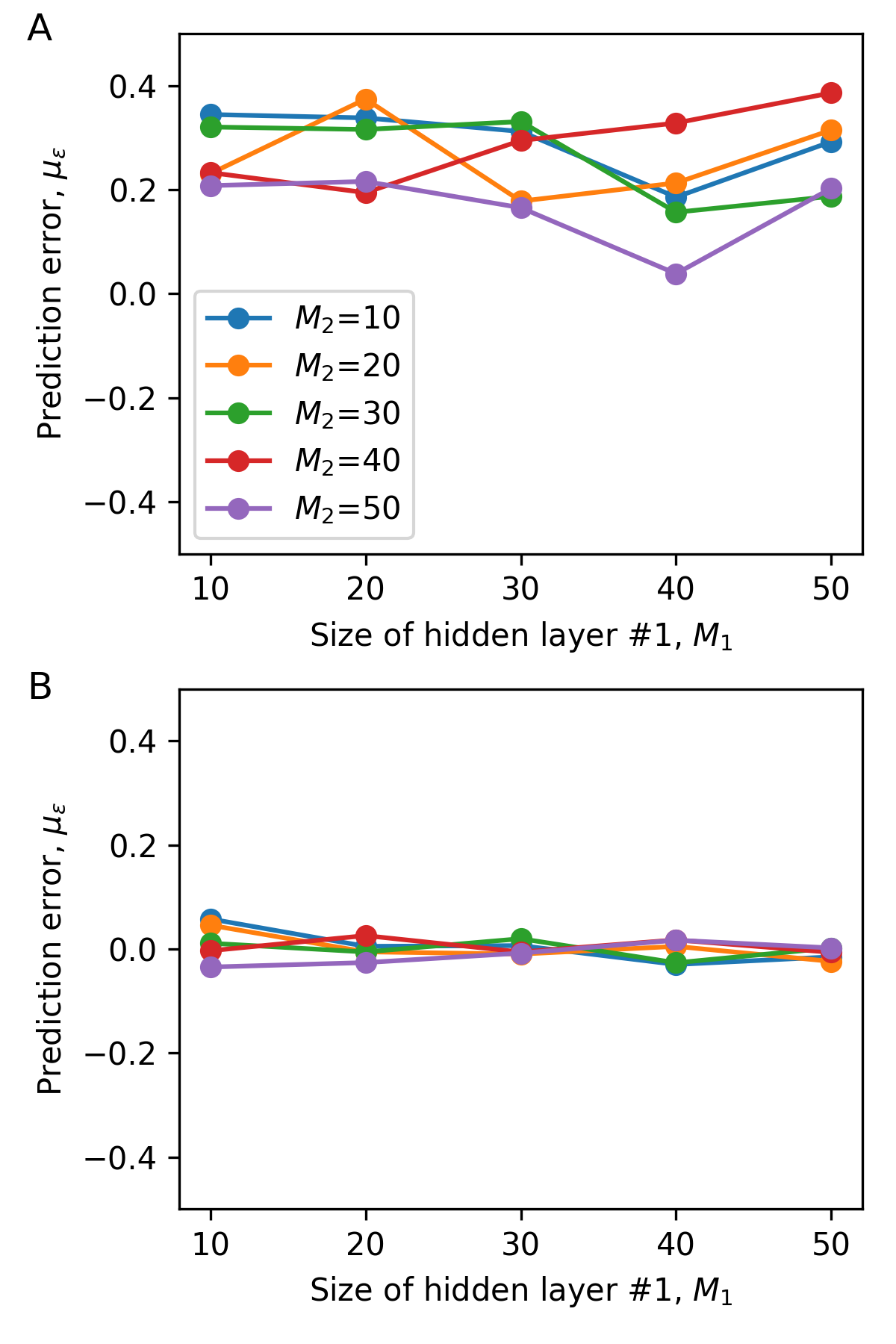}
\caption{\label{fig3} Average prediction error for a surrogate model trained on (A) a single growth profile and (B) two growth profiles. Results are shown as a function of the size of the two hidden layers, $M_1$ and $M_2$ of the artificial neural network used to build the surrogate model. Results are the average of three separate training runs.}
\end{figure}

In Figure \ref{fig3} we compare the mean prediction error,
$\mu_\varepsilon$, of the normalized saturation time $\kappa$ from two different surrogate models: a model trained on thickness data from a single experiment [Fig. \ref{fig3}(A)] and a model trained on the data from two experiments corresponding to two exposures [Fig. \ref{fig3}(B)]. The results have been obtained for the denser dataset, comprising 20 normalized thickness values along the vertical feature, and the corresponding $\mu_\varepsilon$ are shown as a function of the sizes of the hidden layers in our neural network, $M_1$, and $M_2$. The results clearly show that the information contained in a single cross section is not enough to accurately predict the normalized saturation time, resulting in an average overshoot of 25\% with respect to the ground truth saturation time in our dataset. In contrast, two separate experiments do contain enough information to reduce the average error to well below 5\% with the results centered around zero [Fig. \ref{fig3}(B)].

\begin{figure}
\includegraphics[width=7.5cm]{ 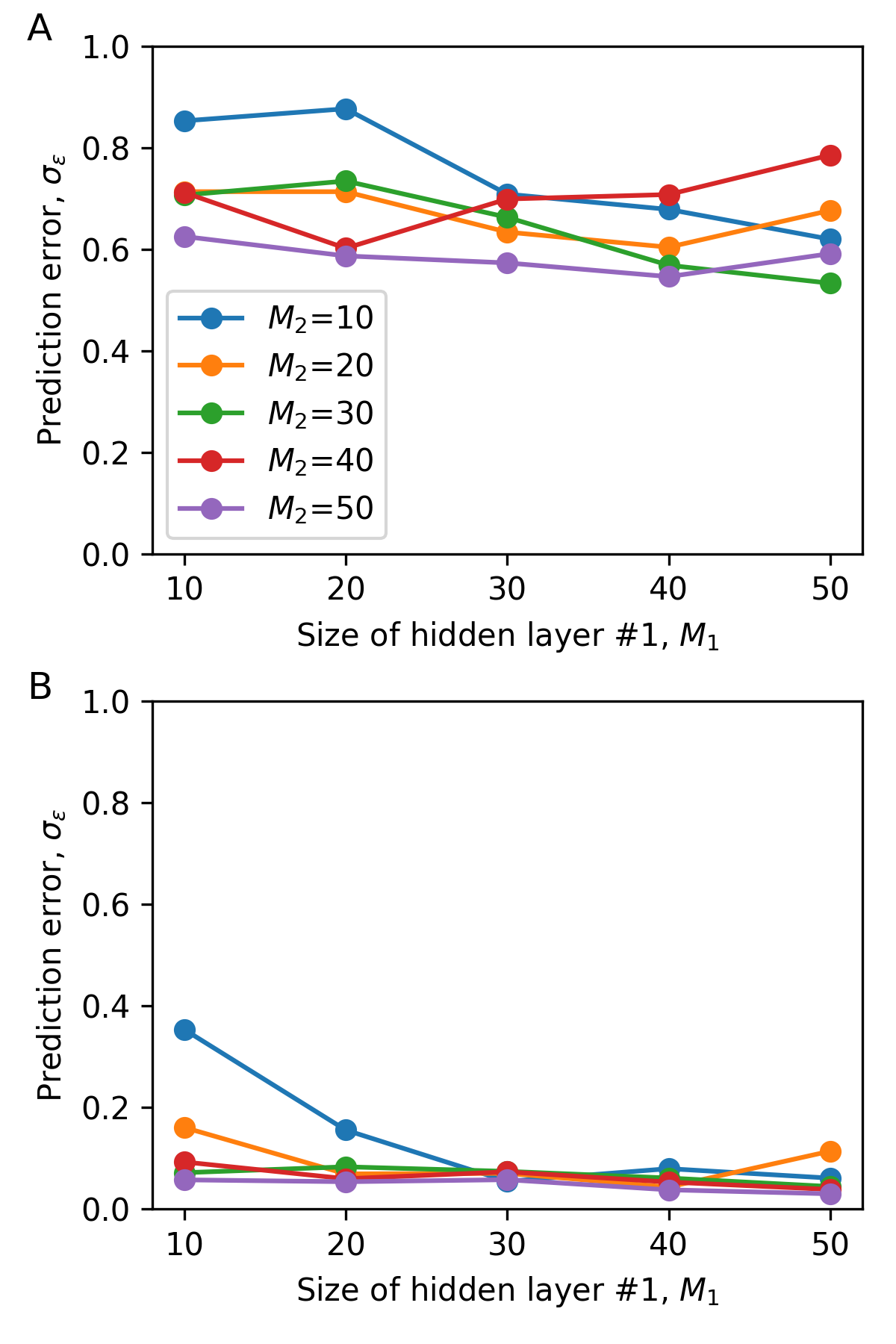}
\caption{\label{fig4} Standard deviation of the predicted saturation times for  a surrogate model trained on (A) a single growth profile and (B) two growth profiles. Results are shown as a function of the size of the two hidden layers, $M_1$ and $M_2$ of the artificial neural network used to build the surrogate model. Results are the average of three separate training runs.}
\end{figure}

In addition to a higher average error, the predictions of the surrogate model based on a single experiment also have a significant statistical deviation (Figure 4(a)). This means that, even though the average error is 25\%, the individual deviations with respect to this value are substantial, and on the order of 70\%. In contrast, the surrogate model trained for two profiles presents a significantly smaller statistical deviation, indicating not only that the model achieves good average performance, but that this performance is close to the ground truth for a majority of the cases in the testing dataset. As the size of the hidden layers increases, the standard deviation reaches a value of around 5\%. Only in the cases where the network size is too small (10 neurons per layer), does
the statistical deviation exceed 20\%. 

\begin{figure}
\includegraphics[width=7.5cm]{ 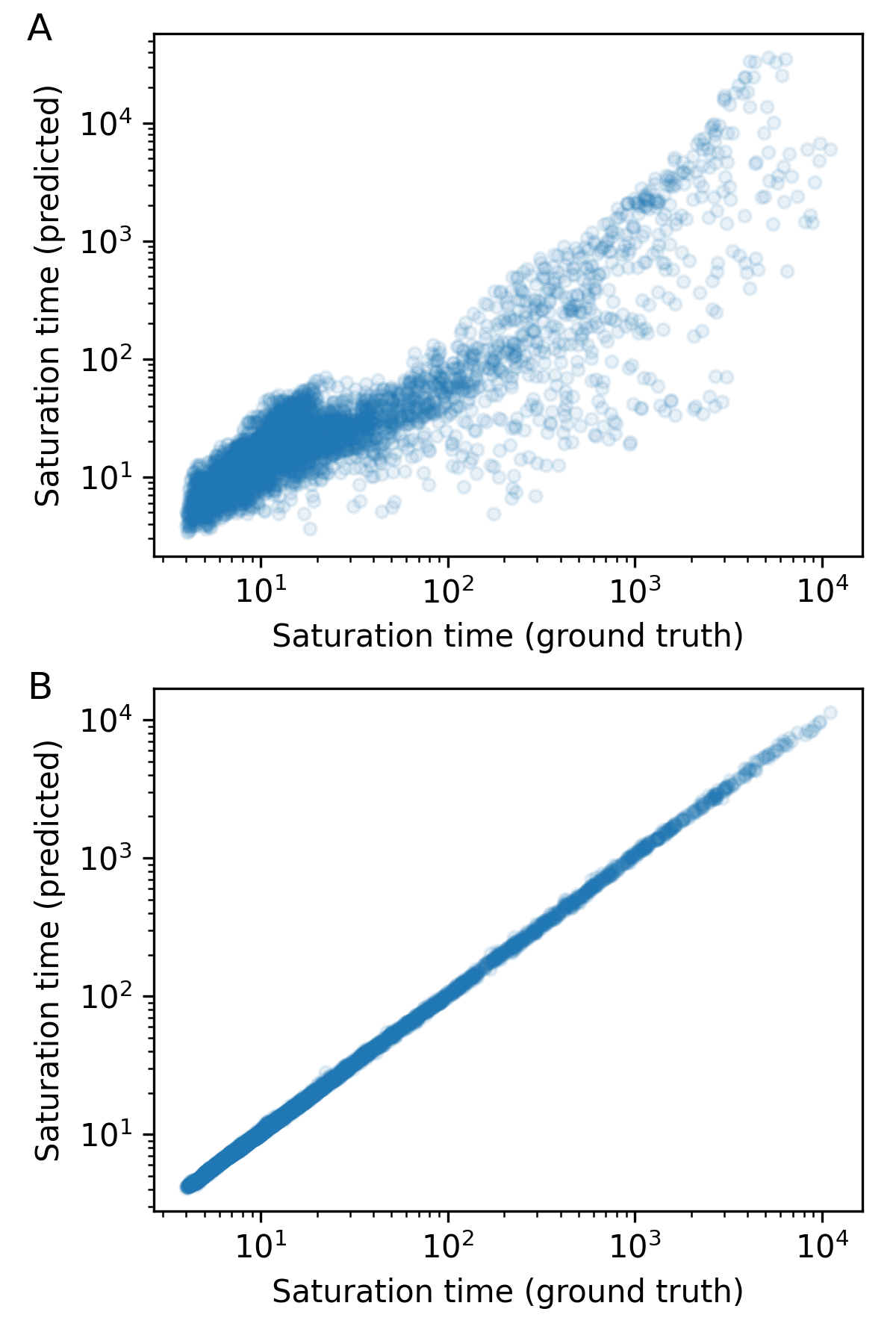}
\caption{\label{fig5} Correlation between the predicted and ground truth normalized saturation times $\kappa$ for surrogate models trained on (A) a single growth profile (B) two growth profiles obtained at different dose times. The size of the hidden layers are $M_1=30$ and $M_2=30$ in both cases.}
\end{figure}

In Figure \ref{fig5} we show the correlation between the predicted normalized saturation time and the ground truth values for surrogate models based on one and two experiments. The plots in Figure \ref{fig5} correspond to the prediction of a surrogate model with hidden layers of size $M_1=30$ and $M_2=30$. The smaller average error and standard deviation when two experiments are used as input to the surrogate model [Fig. \ref{fig5}(B)] translates into much better correlations compared to the single experiment case [Fig \ref{fig5}(A)]. In contrast to the results obtained for the case of growth profiles in thermal ALD\cite{YanguasGil_ALDML_2022}, where saturation times could be predicted using only one experiment, the presence of two separate reaction pathways requires at least two experiments to discriminate the contributions of both the self-limited process and the recombination probability of plasma species to the coating of high aspect ratio features.

\begin{figure}
\includegraphics[width=7.5cm]{ 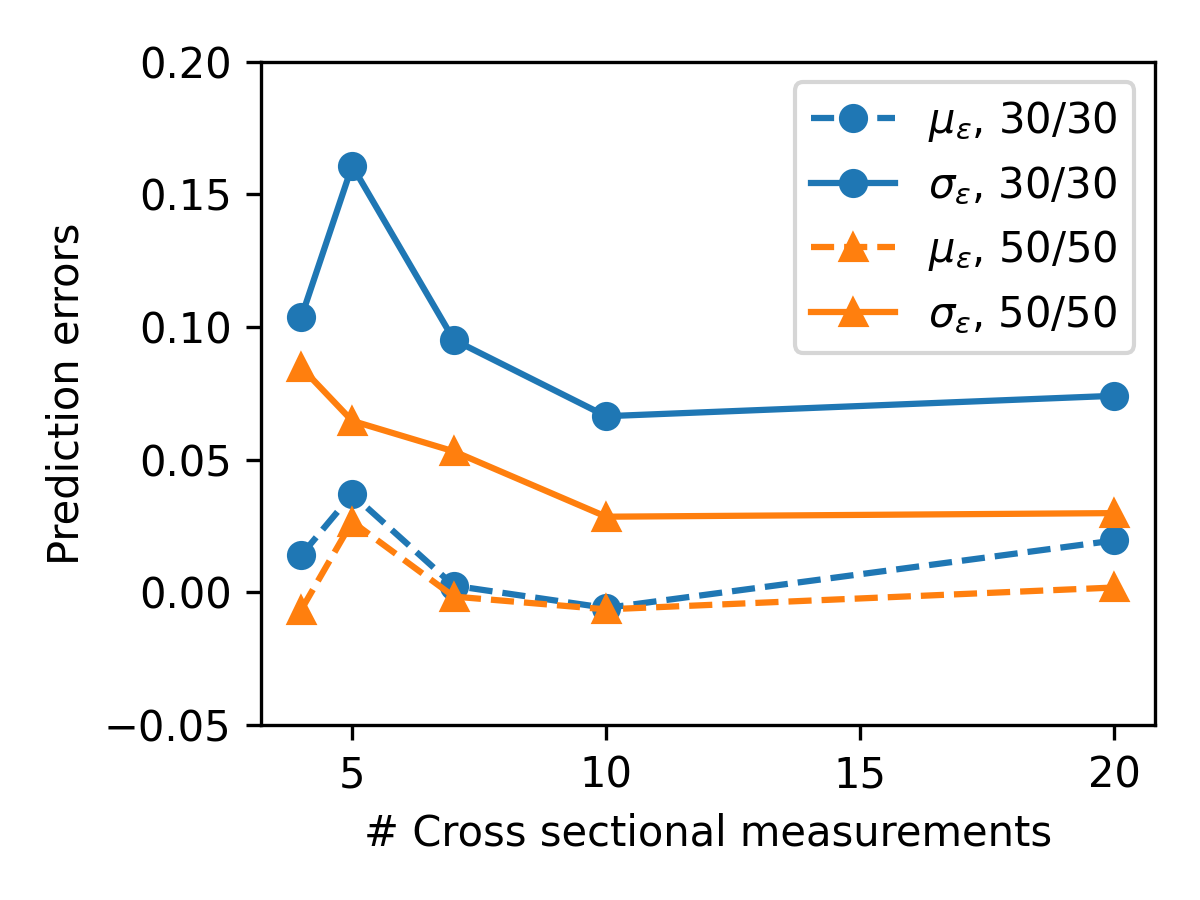}
\caption{\label{fig6} Average prediction error and standard deviation as a function of the number of points in each cross sectional measurement used to train the surrogate model. Results are shown for two networks, one of size $M_1=30$, $M_2=30$, and another of $M_1=50$, $M_2=50$.}
\end{figure}

The results thus far are based on 20 different thickness measurements at different points of the via. To understand how many measurements are required, we built sparser datasets containing fewer measurements along the via. To this end, we took our dataset and we selected 1 out of $n=$ 2, 3, 4, and 5 points along the high aspect ratio via, resulting in thinned datasets comprising 10, 7, 5, and 4 measurements per experiment, respectively. We then trained similar networks with two hidden layers using the same procedure we used for the denser dataset.

\begin{figure}
\includegraphics[width=7.5cm]{ 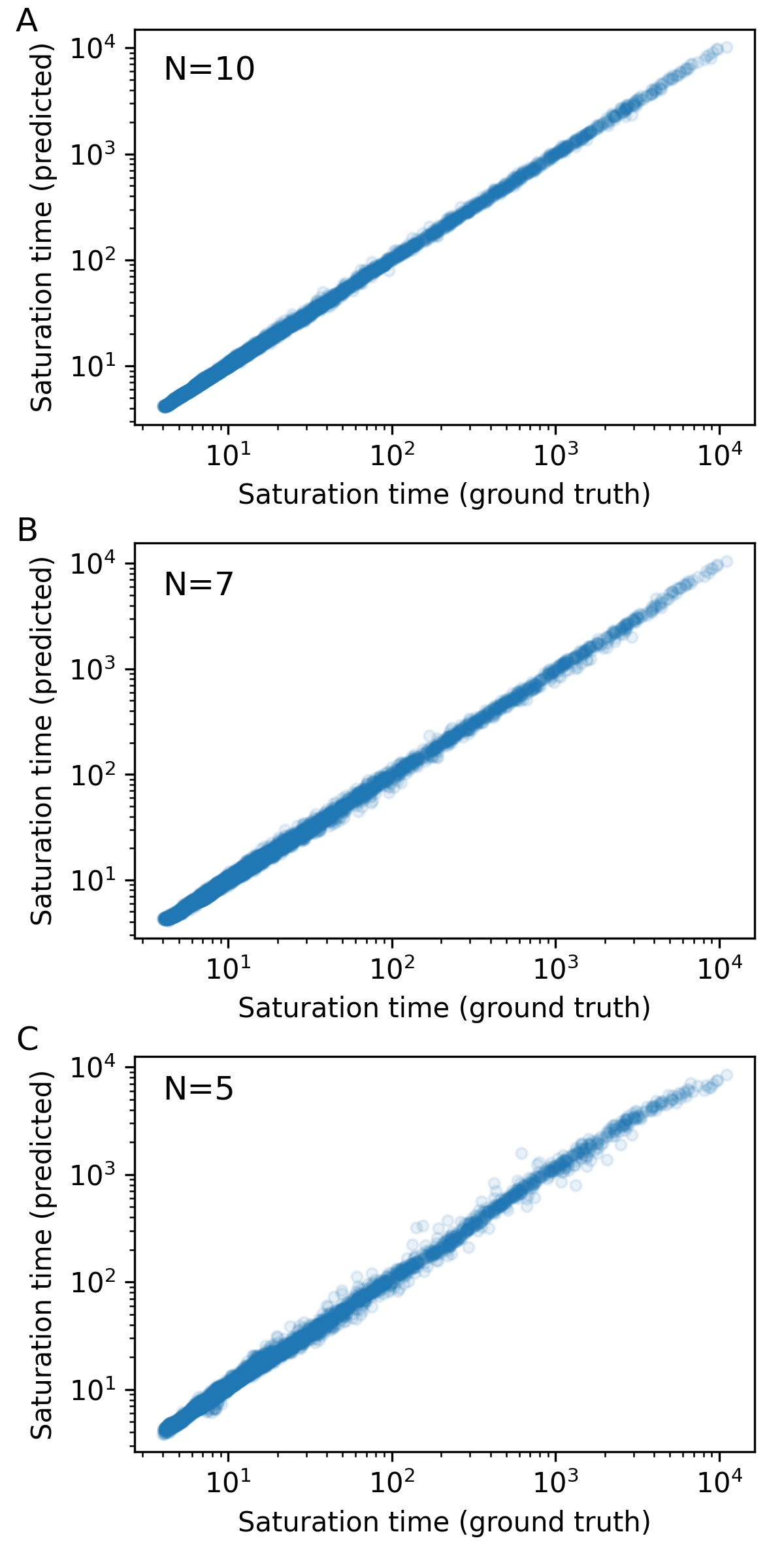}
\caption{\label{fig7} Correlation between the predicted normalized saturation times and the ground truth values for surrogate models trained with different number of points $N$ per cross section measurements. The results show that at least $N=7$ points are needed to minimize the dispersion. In all cases the size of the hidden layers in the artificial neural network are $M_1=30$ and $M_2=30$ neurons.}
\end{figure}

In Figure \ref{fig6} we show the mean prediction error and its corresponding standard deviation for two different networks, one comprising two hidden layers with $M_1=30$ and $M_2=30$ neurons and a second one with $M_1=50$ and $M_2=50$. These results are based on two growth profiles. The results show that as few as 7 points are enough to predict the saturation time with high accuracy. This sets the minimum number of thickness measurements per growth profile required to optimize a plasma-ALD process. In Figure \ref{fig7} we show the correlation between the predicted and ground truth normalized times for $N=10$, $N=7$, and $N=5$ independent points.  The loss in predictive value of the surrogate model is apparent in the increasing spread between the predicted values and the ground truth for $N=5$ [Figure \ref{fig7}(C)].

\begin{figure}
\includegraphics[width=7.5cm]{ 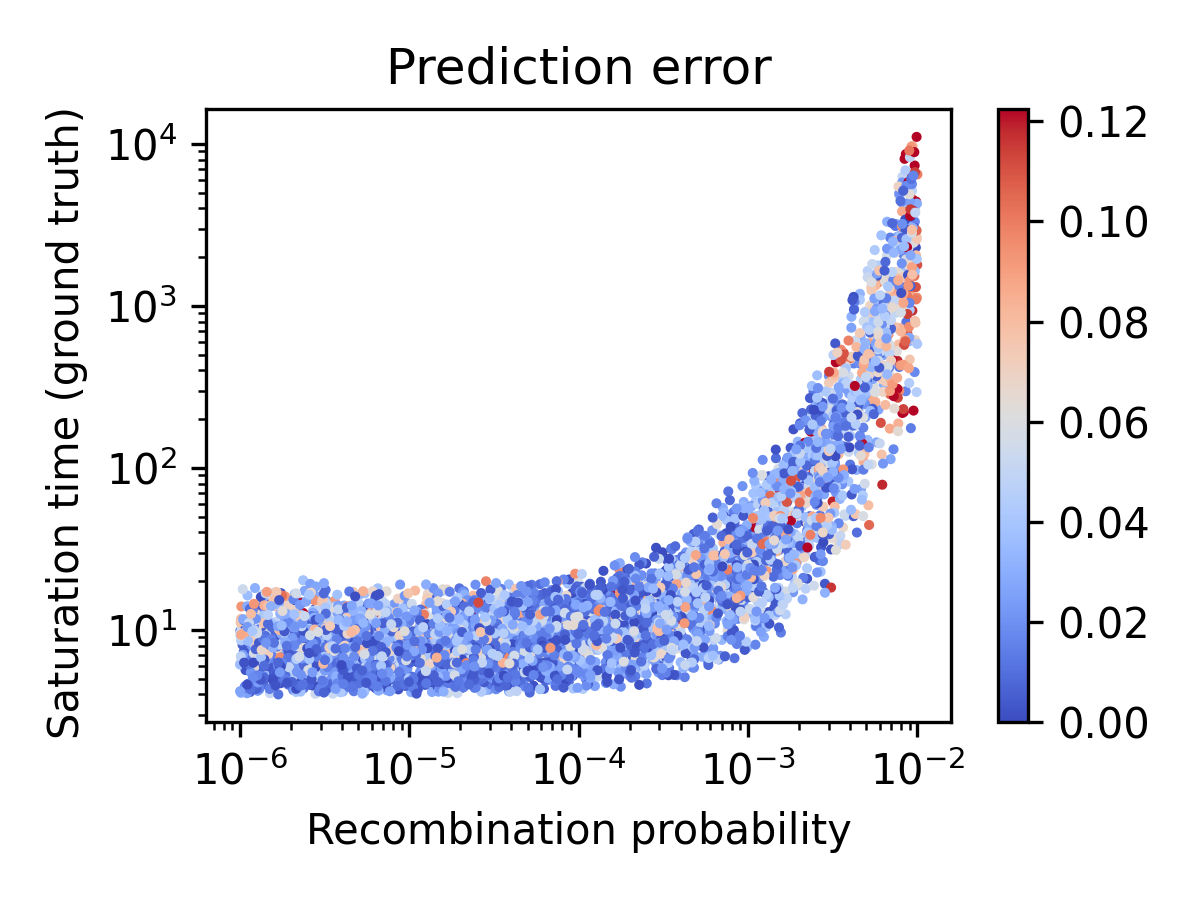}
\caption{\label{fig8}  Prediction errors for each data point in the testing dataset plotted as a function of the recombination probability and the normalized saturation
time of each data point. The results are shown for datasets involving $N=20$ points, and a network with $M_1=50$, and $M_2=50$ neurons.}
\end{figure}

The results in Figures \ref{fig3} - \ref{fig7} establish that it is possible to predict saturation times based on two separate profiles with low errors. However, even
in the largest network considered in this work, comprising 50 hidden neurons in each layer, the standard deviation of the surrogate model is still of the order of 4\%. 
To ascertain if errors are concentrated on specific regions (i.e. high sticking or recombination probabilities) that
are somehow not well represented in the dataset, in Figure \ref{fig8} we represent the prediction error
for each point in our testing set as a function of two parameters: the recombination probability and the predicted normalized saturation time. The results
are shown for $N=20$, $M_1=50$, and $M_2=50$. While we observe that a concentration of points with high prediction errors are concentrated for high recombination
probabilities, this does not fully explain the presence of lingering errors in the surrogate model. It is more likely that the remaining error is therefore due to the
limited expressivity of the network sizes considered in this work.

\subsection{Extracting surface kinetics information}

The surrogate models thus far have focused on predicting optimal process conditions. However, one alternative approach would be to use profile data to extract the values of the sticking probability for the self-limiting process and the recombination probability for the atomic species used in the PEALD process. This data can in turn be fed into simulations to predict the saturation times. In order to evaluate the feasibility of this approach we have used the same methodology as for the normalized saturation times and trained artificial neural networks to predict the sticking and recombination probabilities. 

\begin{figure}
\includegraphics[width=7.5cm]{ 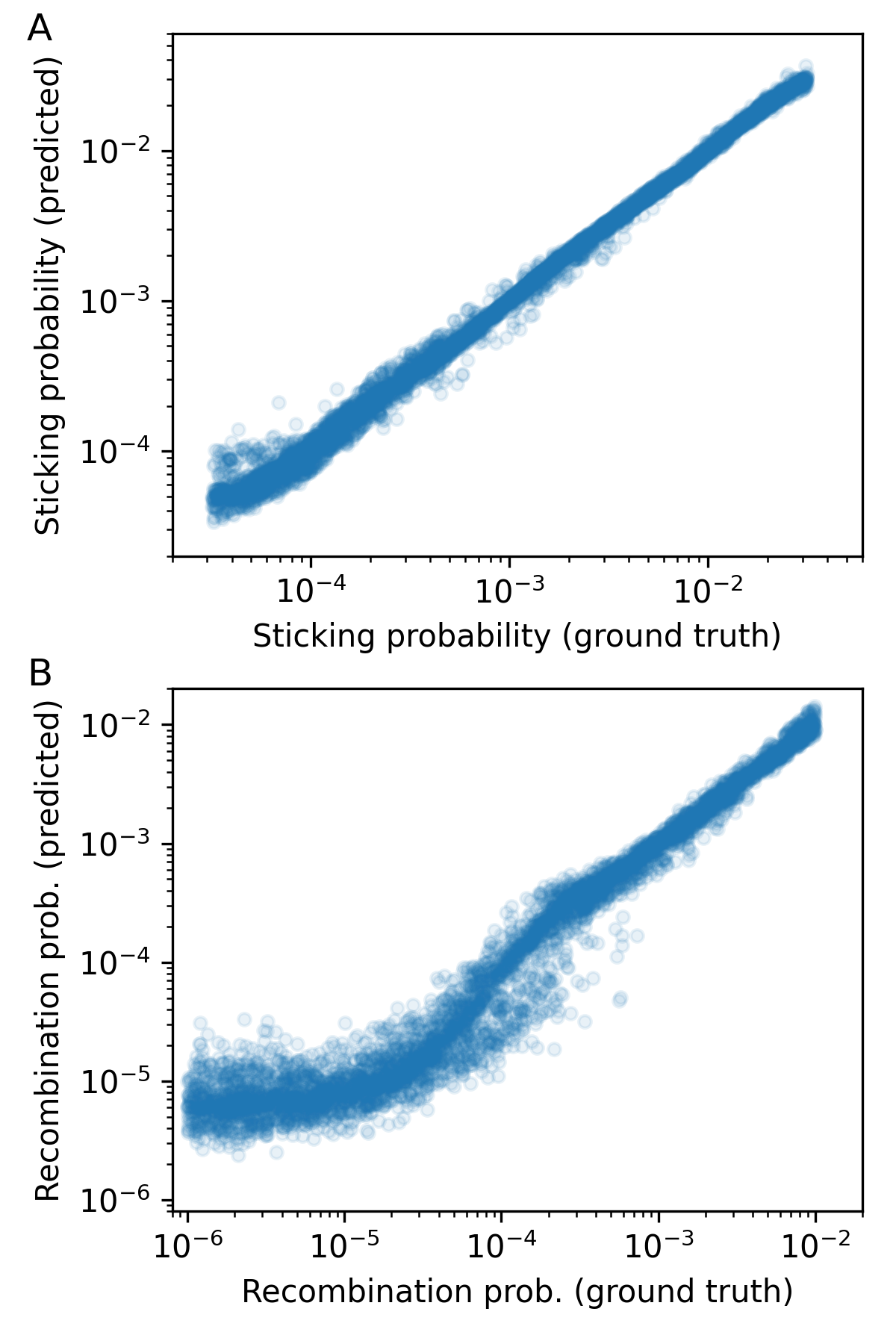}
\caption{\label{fig9} Correlation between predicted and ground truth values for (A) the sticking probability and (B) recombination probability from surrogate models trained on two separate growth profiles. The sizes of the hidden layers of the artificial neural network are $M_1=30$, $M_2=30$}
\end{figure}

In Figure \ref{fig9} we show the correlation between the predicted and ground truth values for the sticking and recombination probabilities in our testing dataset. In both cases, the surrogate model is trained on the dense dataset comprising two profiles with 20 different points per condition. Compared to the predictions for the normalized saturation time, the dispersion in the data is significantly larger. This indicates that it is harder to extract the kinetic data for the individual reaction pathways than it is to predict the saturation value. In contrast, when we reduce the task to determining whether the surface kinetics is dominated by the self-limited interaction or the surface recombination, the results show excellent agreement with the ground truth.

\begin{figure}
\includegraphics[width=7.5cm]{ 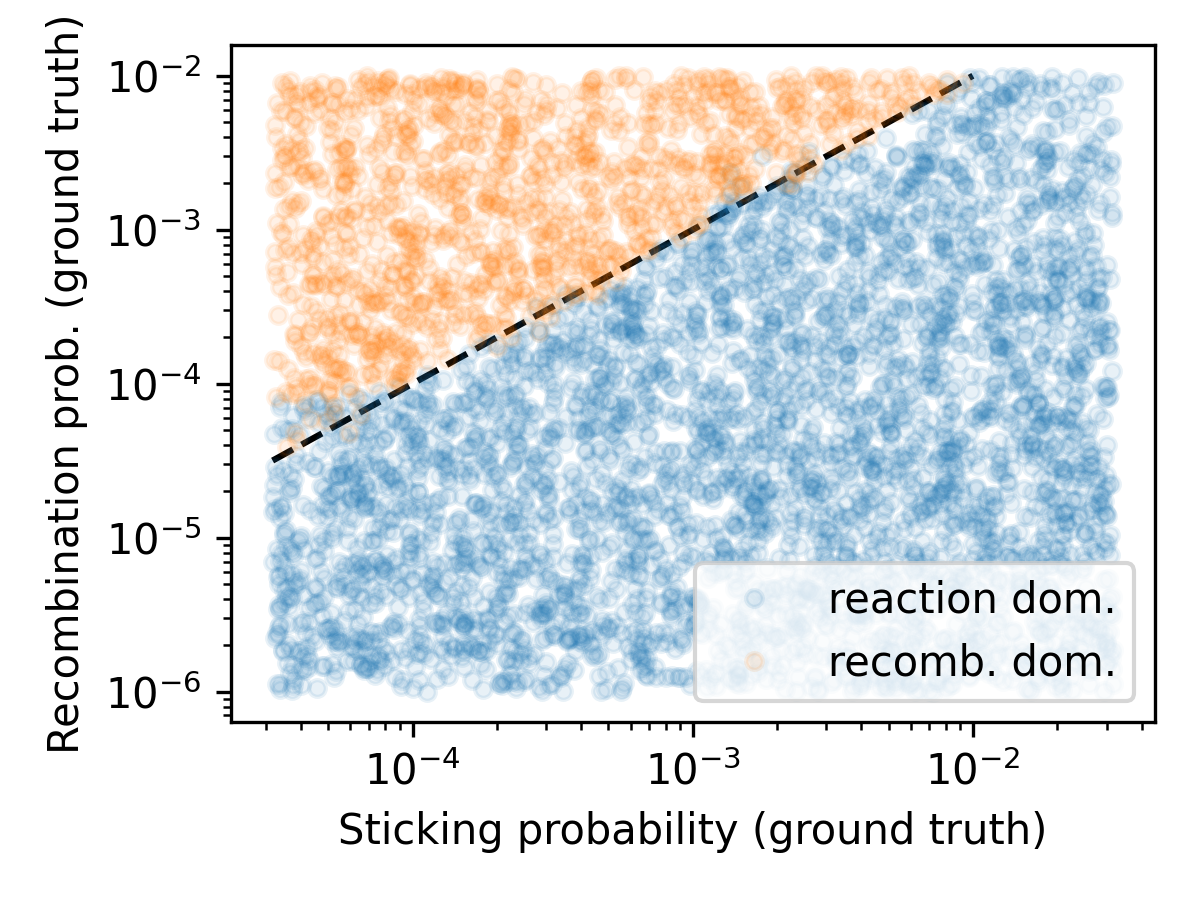}
\caption{\label{fig10} Ability of a surrogate model to discriminate between a reaction dominated ($\beta_s > \beta_r$) and a recombination dominated
($\beta_r > \beta_s$) process. Data points are color coded depending on whether the model classifies the process as a reaction or recombination dominated process. The dash line represents the ideal demarcation between the two regimes}
\end{figure}

In Figure \ref{fig10} we have plotted each sample in the testing dataset as a function of its reaction and recombination probabilities. Each point is then color coded depending on whether the model assigns it as a reaction or recombination dominated process. The dashed line in Figure
\ref{fig10} represents the boundary separating these two regimes. This corresponds to the $\beta_s = \beta_r$ condition. From this plot, it is clear that the model correctly assigns the process to the right category, with an overall classification accuracy exceeding 99\%. Processes with very low sticking and recombination probabilities are the sole exception. This is expected as in these cases the growth profiles appear conformal even at low coverages.

\section{Discussion}

The results obtained in Section \ref{sec:results} conclusively show that the information contained in growth profiles can be used to predict optimal dose times on plasma-assisted self-limited processes where the reaction pathways leading to thin film growth must compete with surface recombination pathways. These are ubiquitous in radical species such as atomic hydrogen, atomic oxygen, or atomic nitrogen. This is something that can be leveraged to accelerate the evaluation or optimization of a novel process for which kinetic data is not well understood. This method could also be applied to thermal chemistries that experience surface deactivation pathways such as metal oxide ALD using
ozone\cite{Knoops_ozone_2011}.

In this work we have focused on a relatively simple case where the ALD process is dominated by a single plasma species. We have also considered the case of a circular via with a fixed aspect ratio. A key question is therefore how generalizable the results are. There are several areas where the generalization seems, in principle, trivial: for instance, while we focused on the case of ALD, the extension to atomic layer etching seems trivial as long as etching and recombination are each dominated by a prevalent reaction pathway. Extension to more complex features is also trivial, as is replacing the Knudsen diffusion model used in this work with more precise ballistic transport models\cite{YanguasGil_ballistic_2014}. In all these cases, the complexity does not increase the number of degree of freedoms in the underlying kinetics, and the methodologies explored in this work are likely to carry over for large enough training datasets.

Challenges therefore come whenever the assumptions in the model are too simple to capture the complexity of the experimental system. Here we would like to highlight two different scenarios: first, the presence of reactor effects, where upstream depletion can significantly reduce the local flux of reactive species, may necessitate more complex models that couple reactor scale and local transport. One way of overcoming this challenge is through the collection of growth profiles at different points in the reactor. For thermal ALD processes, we have observed that information on thickness inhomogeneity at a reactor scale can be used to correctly predict saturation times with the information contained in a single experiment\cite{YanguasGil_ALDML_2022}. The second scenario involves processes driven by multiple surface reaction pathways or plasma species, such as ions and radicals\cite{ Arts_ionconformality_2021}. This scenario may require more than two experiments to accurately predict optimal conditions.

Even in these more complex situations, simpler models can offer insights into the underlying surface kinetics. For instance, large departures from predicted saturation times or the measurement of growth profiles that are not captured in a dataset constructed based on simple assumptions would signal the presence of a more complex surface kinetics that is not contained in the training data. As an example, experimental growth profiles by Arts et al showed a persistent slope that cannot be explained by the models used in this work\cite{Arts_TMAsticking_2019}. This is consistent with the presence of additional phenomena not included in our simulations.

With these considerations, the development of a diagnostic tool based on the analysis of growth profiles would benefit from a few generalizations. These include 
incorporating different aspect ratios as well as expanding the model to other types of structures, such as tapered vias, of relevance in semiconductor processing. Likewise, it should be possible to generalize the model so that it can operate with two arbitrary dose times. Developing a diagnostic tool would also require experimental validation of the model predictions. This is something that we were not able to do for the current model based on the data available in the literature. There is
also an opportunity to explore other methodologies, such as Gaussian process regression, that can quantify the uncertainty of the model predictions. 

Finally, it is important to emphasize the small cost of training these surrogate models. The networks considered here are many orders of magnitude smaller than traditional image classification networks, let alone large language models. The feasibility of studying much more complex systems is therefore not limited by the cost of training, but by the cost of running the simulations or models used to generate the dataset. Also, the generation of these datasets does not require model validation for each specific process. Instead, we need models that are physically correct and can capture the diversity of all possible processes so that surrogate models can learn the correlations between experimental observables used as inputs and the desired model output. There is therefore an opportunity to revisit existing plasma and plasma-surface interaction models to generate datasets that can help us acquire a better understanding of the fundamentals of plasma-surface interactions in different plasma-based processes. As shown in this work, besides process optimization, surrogate models can be used to design experimental protocols that maximize the amount of information obtained while minimizing the number of experiments.

\section{Conclusions}

Our results show that cross-section images from a few experiments contain enough information to predict optimal saturation times for self-limited processes in high aspect ratio features in the presence of surface recombination pathways of plasma species. For the simplest case considered here, involving one dominant self-limited reaction and one dominant recombination pathway, as few as two experiments are enough to predict saturation times with great accuracy (5\% error). This strongly suggests that machine learning can enable new methods to accelerate the optimization of PEALD processes in areas such as microelectronics.

\begin{acknowledgments}
This research is based on work supported by Laboratory Directed Research and Development (LDRD) funding from Argonne National Laboratory, provided by the Director, Office of Science, of the U.S. DOE under Contract No. DE-AC02-06CH11357.
\end{acknowledgments}

\section*{Author declarations}

\subsection*{Conflict of Interest}

The authors have no conflicts to disclose.

\section*{Data Availability Statement}

The data that support the findings of this study has been made openly available in github: \href{https://github.com/aldsim/plasmaml}{https://github.com/aldsim/plasmaml}

\bibliography{aipsamp}

\end{document}